\documentclass[12 pt,leqno]{article}%
\usepackage{latexsym}
\usepackage{amssymb,amsbsy,amsmath,amsfonts,amscd}
\usepackage{vmargin}
\usepackage{epsfig}
\usepackage{amsmath}
\usepackage{amsfonts}
\usepackage{amssymb}
\usepackage{graphicx}%
\providecommand{\U}[1]{\protect\rule{.1in}{.1in}}

\newtheorem{theorem}{Theorem}[section]
\newtheorem{definition}{Definition}[section]
\newtheorem{proposition}{Proposition}[section]

\newcommand{\N}{\ifmmode{{\rm I} \hskip -2pt {\rm N}}
    \else{\hbox{$I\hskip -2pt N$}}\fi}
\newcommand{\R}{\ifmmode{{\rm I} \hskip -2pt {\rm R}}
    \else{\hbox{$I\hskip -2pt R$}}\fi}

\newcommand{ \vit}{\hbox{\bf u}}

\newcommand{ \wit}{\hbox{\bf w}}

\newcommand{\x} {{\bf x}}
\newcommand{\BEQ} {\begin{equation} }
\newcommand{\EEQ} {\end{equation} }
\makeatletter
\@addtoreset{equation}{section}

\begin{document}

\title{Attractors for a  deconvolution model of turbulence}
\author{Roger Lewandowski\thanks{IRMAR, UMR 6625,
Universit\'e Rennes 1,
Campus Beaulieu,
35042 Rennes cedex
FRANCE; 
Roger.Lewandowski@univ-rennes1.fr, 
http://perso.univ-rennes1.fr/roger.lewandowski/} and Yves Preaux\thanks{Lyc\'ee du Puy de L\^ome, rue du Puy de L\^ome, 29 200 Brest, Yves.Preaux@free.fr}}

\maketitle

\begin{abstract}

We consider a deconvolution model for 3D periodic flows. We show the existence of a global attractor for the model. 

\end{abstract}

MCS Classification : 76D05, 35Q30, 76F65, 76D03
\medskip


\section{Introduction}

This note is concerned by the deconvolution model of order $N$ introduced in \cite{LL08} (model $(\ref{pb})$ below) for 3D periodic flows. This model takes inspiration in the class of the so called $\alpha$-models (see in \cite{CHOT05} and  \cite{Fo01} and references inside) and also in the class of ADM models (see in \cite{St01}). We are interested by the question of the existence of a global  attractor for this model. 
\smallskip

The question of attractors has already been considered for the alpha model (see 
\cite{CTV07}), corresponding to the case $N=0$. We prove in this work the existence of an attractor for each $N$ (see Theorem \ref{MAIN}). 
\smallskip

In order to make the paper self contained, we describe carrefully how is constructed the 
deconvolution model. Next, we recall basic notions on the attractors, notions that can be founded in the book of R. Temam (see \cite{RT88}). Finally we prove the existence of the attractor. The question of its dimension is under progress. 

\section{The Deconvolution model }

\subsection{Function Spaces} 

for $s \in\ifmmode{{\rm I} \hskip -2pt {\rm R}}
\else{\hbox{$I\hskip -2pt R$}}\fi$, let us define the space function
\begin{equation} \begin{array} [c]{l}Ê
\mathbf{H}_{s} = \left\{  \mathbf{w}  =\sum_{\mathbf{k}}\widehat{\wit} e^{i {\bf k} \cdot {\bf x}},  \, \, \nabla\cdot\mathbf{w} = 0, \, \, \widehat\wit(
\mathbf{0} )= \mathbf{0}, \quad\sum_{\mathbf{k} } | \mathbf{k} |^{2s}
|\widehat{\wit }(\mathbf{k},t)|^{2} < \infty\right\}  . \end{array} 
\end{equation}
We define the $\mathbf{H}_{s}$ norms by
\begin{equation}
|| \wit ||^{2}_{s} = \sum_{\mathbf{k} } | \mathbf{k} |^{2s} |\widehat
{\wit }(\mathbf{k},t)|^{2},
\end{equation}
where of course $|| \wit ||^{2}_{0} = || \wit ||^{2}$. It can be shown that
when $s$ is an integer, $|| \wit ||^{2}_{s} = || \nabla^{s} \wit ||^{2}$ (see
\cite{DG95}).

$\newline$
We denote by $P_{L}$ The Helmholtz-Leray orthogonal projection of $(L^{2})^{3}$ onto $\mathbf{H}_{0}$ and by $A$ the Stokes operator defined by $A=-P_{L}\triangle$ on $D(A)=\mathbf{H}_{0}\cap(H^{2})^{3}$. We note that in the space-periodic case  $A\mathbf{w}=-\triangle\mathbf{w}$ for all  $\wit\in D(A)$.\\
The operator $A^{-1}$ is a sef-adjoint positive definite compact operator from $\mathbf{H}_{s}$ onto $\mathbf{H}_{s}$, for $s=1$ and $s=2$ (see \cite{FMRT01}). We denote $\lambda_{1}$ the smallest eigenvalue of $A$.\\
We introduce the trilinear form b defined by
\begin{equation}
b(\vit,\mathbf{v},\wit) = \sum_{\mathbf{i,j} }\int_{\Omega
} u_{i}\partial_{i}v_{j}w_{j}dx.
\end{equation}
wherever the integrals make sense.
Note that $b(\vit,\wit,\wit) =0$ when $\nabla\cdot\vit=0$.

\subsection{The Filter and the deconvolution process}Ê

Let $\wit\in\mathbf{H}_{0}$ and $\overline{\wit}\in\mathbf{H}_{1}$ be the
unique solution to the following Stokes problem with periodic boundary conditions:
\begin{equation}
-\delta^{2}\triangle\overline{\wit}+\overline{\wit}+\nabla r=\wit\quad\text{
in }\mathbb{R}^{3},\quad\nabla\cdot\overline{\wit}=0,\quad\int_{\Omega
}\overline{\wit}=\mathbf{0}. \label{filter}%
\end{equation}
We denote the filtering operation by $G$
so that $\overline{\mathbf{w}}=G\mathbf{w}$. Writing
$
\wit(\mathbf{x},t)=\sum_{\mathbf{k}}\widehat{\wit}(\mathbf{k}%
,t)e^{-i\mathbf{k \cdot  x}},
$
it is easily seen that $\nabla r=0$ and
$\displaystyle 
\overline{\wit}(\mathbf{x},t)=\sum_{\mathbf{k}}{\frac{\widehat{\wit}%
(\mathbf{k},t)}{1+\delta^{2}|\mathbf{k}|^{2}}}e^{-i\mathbf{k \cdot  x}}.%
$
Then writing $\overline{\wit}=G(\wit)$, we see that in the corresponding
spaces of the type $\mathbf{H}_{s}$, the transfer function of $G$, denoted by
$\widehat{G}$, is the function
$
\widehat{G}(\mathbf{k})={\frac{1}{1+\delta^{2}|\mathbf{k}|^{2}}},
$
and we also can write on the $\mathbf{H}_{s}$ type spaces 
\begin{equation}
-\delta^{2}\triangle\overline{\wit}+\overline{\wit}=\wit\quad\text{ in
}\mathbb{R}^{3},\quad\nabla\cdot\overline{\wit}=0,\quad\int_{\Omega}%
\overline{\wit}=\mathbf{0}.
\end{equation}

The procedure of deconvolution by the Van Citter approximation is described in \cite{LL08}.  This yields the operator 
$
D_{N}\wit=\sum_{n=0}^{N}(I-G)^{n}\wit.
$
\begin{definition}
The truncation operator $H_{N}: \mathbf{H}_{s} \rightarrow\mathbf{H}_{s} $ is
defined by 
$
H_{N}\wit:=D_{N}\overline{\wit }=(D_{N}\circ G)\wit .
$ \hfill $\blacksquare$
\end{definition}Ê
Note that, for any $s \geqslant 0$ we have the following proprieties (see \cite{LL08}) :
\begin{equation}
\left\Vert H_{N}\wit\right\Vert_{s}\leqslant \left\Vert \wit\right\Vert_{s},
\label{H2}%
 \quad \left\Vert H_{N}\wit\right\Vert_{s+2}\leqslant C( \delta,N)\left\Vert \wit\right\Vert_{s}.
\end{equation}
\subsection{The model}Ê
\vskip -0,3cm

Let $\vit_{0}\in\mathbf{H}_{0}$, $f\in\mathbf{H}_{-1}$. For $\delta>0$, let
the averaging be defined by $(\ref{filter})$. The problem we consider is the
following: for a fixed $T>0$, find $(\wit, q)$
\begin{equation}
\label{pb}\left\{
\begin{array}
[c]{l}%
\wit \in L^{2} ([0,T], \mathbf{H}_{1}) \cap L^{\infty}([0,T], \mathbf{H}_{0}),
\quad\partial_{t}\wit \in L^{2} ([0,T], \mathbf{H}_{-1})\\
q \in L^{2} ([0,T], L^{2}_{ {\hbox{\footnotesize per}, 0} }),\\
\partial_{t}\wit+ (H_{N} (\wit) \cdot \nabla)\,  \wit-\nu\triangle\wit+\nabla
q=H_{N}(\mathbf{f}) \quad\text{ in } \mathbb{\mathcal{D}}^{\prime}( [0,T]
\times\ifmmode{{\rm I} \hskip -2pt {\rm R}}
\else{\hbox{$I\hskip -2pt R$}}\fi^{3}),\\
\wit ( \mathbf{x}, 0) = H_{N} (\vit_{0}) = \wit_{0}.
\end{array}
\right.
\end{equation}
where $L^{2}_{{\hbox{\footnotesize per}, 0} }$ denotes the scalar fields in
$L^{2}_{loc} (\ifmmode{{\rm I} \hskip -2pt {\rm R}}
\else{\hbox{$I\hskip -2pt R$}}\fi^{3})$, $2 \pi$-periodic with zero mean value.
We prove in \cite{LL08}Ê the following result. 
\begin{theorem}
Problem $(\ref{pb})$ admits a unique solution $(\wit, q)$, $\wit \in L^{\infty}([0,T], \mathbf{H}_{1}) \cap L ^{2} ([0,T],
\mathbf{H}_{2})$, and 
the following energy equality holds:
\begin{equation}
\label{ENER}\frac{1}{2}\left\Vert \wit (t)\right\Vert ^{2}+ \nu\int_{0}%
^{t}\int_{\Omega}|\nabla\wit|^{2}d\x dt^{\prime}=\frac{1}{2}\left\Vert H_{N}
(\vit_{0})\right\Vert ^{2}+\int_{0}^{t}\int_{\Omega} H_{N} (\mathbf{f}) .
\wit \, d\x dt^{\prime}. \hskip 0.5cm \blacksquare
\end{equation}
\end{theorem}

\section{Main result} \vskip -0,1cm
\subsection{Recall of basic notions about attractors}Ê
\vskip -0,2cm
We denote by $ \wit (t, \cdot) = S(t) (\wit_0)$ the (unique) solution of system 
$(\ref{pb})$ at time $t$. We recall the definitions of  a global attractor and an absorbing set (see in 
\cite{RT88}). 

\begin{definition} 
We say that ${\cal A} \subset {\bf H}_0$ is a global attractor for the dynamical system 
$(\ref{pb})$ if and only if 
\smallskip

(P1) ${\cal A}$ is compact in the space ${\bf H}_0$,
\smallskip

(P2) $\forall \, t \in \R$, $S(t) ({\cal A}) \subset {\cal A}$,
\smallskip

(P3) For every bounded subset $B \subset {\bf H}_0$, $\rho (S(t)(B), {\cal A})$ goes to zero when $t$ goes to infinity, where 
$ \rho (S(t)(B), {\cal A}) =  \sup_{v \in B}Ê\inf_{u \in {\cal A}} ||Êu - v||.$ \hfill $\blacksquare$
\end{definition}Ê 
\vskip -0,2cm \begin{definition}Ê\label{RRTP}
1. A set ${A}Ê\subset {\bf H}_0$ is an absorbing set if and only if for every bounded subset $B \subset {\bf H}_0$ there exists $t_1 >0$ such that for all $t \ge t_1$ one has 
$S(t) (B) \subset {A}$. 
\smallskip
2. We say that the semi group $S(t)$ is uniformly compact if and only if  for every bounded subset $B \subset {\bf H}_0$ there exists $t_2 = t_2(B) $ such that
$ \displaystyle \overline {{\bigcup_{t \ge t_2} S(t) (B)}}$ is compact.  
\smallskip

3. We denote by  $\omega (A) $ the set $\displaystyle \omega (A) = \bigcap_{s \ge 0} \overline {\bigcup_{t \ge s} S(t) (A)}$.  \hfill $\blacksquare$

\end{definition}Ê

\begin{proposition}Ê\label {ATTR}
Assume that there exists an absorbing bounded set $A$ and that 
the semi group $S(t)$ is uniformly compact, then  ${\cal A}Ê= \omega (A)$ is the global attractor for the dynamical system defined by $S(t)$.  
\end{proposition}Ê 
see the proof in  \cite{RT88}).  
\vskip -0,4cm  \subsection{Existence of a global attractor}Ê
\vskip -0,3cm
We are now in order to state and prove the main result of this note. 

\begin{theorem}Ê\label{MAIN} 
The system $(\ref{pb})$ has a global attractor. \hfill $\blacksquare$
\end{theorem}Ê
\vskip -0,4 cm
{\bf Proof.} Thanks to Proposition \ref{ATTR}, it remains to prove that system Ê
$(\ref{pb})$ has an absorbing set and that $S(t)$ is uniformly compact, in the sense of 
definition \ref{RRTP}. Both things are derived from basic estimates that we detail in the following.  
\smallskip

{\bf Absorbing set in ${\bf H}_0$ : } We take the inner product of  the first Žquation of system $(\ref{pb})$ with $\wit$ to obtain 
\begin{equation}
\frac{1}{2}\frac{d}{dt}\left\Vert \wit\right\Vert ^{2}+ b(H_{N} (\wit), \wit, \wit)+\nu|| \wit ||^{2}_{1}=(H_{N}(\mathbf{f}),\wit).
\end{equation}
%
Observing that $b(H_{N} (\wit), \wit, \wit)=0$ due to $\nabla\cdot H_{N} (\wit)=0$, applying Young inequality, Poincare inequality $|| \wit ||\leqslant\lambda_{1}^{-\frac{1}{2}}|| \wit ||_{1}$ and  using  $(\ref{H2})$ there remains
\begin{equation}\label{ineq2}
\frac{d}{dt}\left\Vert \wit\right\Vert ^{2}+\nu\lambda_{1}|| \wit ||^{2}\leqslant\frac{1}{\nu\lambda_{1}}\left\Vert \mathbf{f}\right\Vert^{2}.
\end{equation}
So, noting $\rho_{0}=\frac{1}{\nu\lambda_{1}}\left\Vert \mathbf{f}\right\Vert$ and applying Gronwall lema we obtain 
\begin{equation}\label {estimationH}
\left\Vert \wit\right\Vert ^{2}\leqslant\left\Vert \wit_{0}\right\Vert ^{2}e^{-\nu\lambda_{1} t}+\rho^2_{0}(1-e^{-\nu\lambda_{1} t}).
\end{equation}
Considering $\wit_{0}$ included in a ball $B(0,R)$ and choosing $\rho_{0}^{\prime}>\rho_{0}$, the previous inequality implies that, for $t>T_{0}$, 
\begin{equation}\label {estimationH}
\left\Vert \wit(t)\right\Vert ^{2}<{\rho_{0}^{\prime}}^2,  \quad with \quad T_{0}=\frac{1}{\nu\lambda_{1}}ln\frac{R^2}{{\rho_{0}^{\prime}}^2-{\rho_{0}}^2}.
\end{equation}
Since each bounded set of  $\mathbf{H}_{0}$ is included in a ball $B(0,R)$, one deduces that $B(0,\rho_{0}^{\prime})$ is an absorbing set in  $\mathbf{H}_{0}$.

$\newline$
More, as an alternative of $(\ref{ineq2})$ we may obtain
\begin{equation}
\frac{d}{dt}\left\Vert \wit\right\Vert ^{2}+\nu|| \wit ||^{2}_{1}\leqslant\frac{1}{\nu\lambda_{1}} \left\Vert \mathbf{f} \right\Vert^{2} .
\end{equation}
Integrating between $t$ and $t+r$, we observe than, for $ \vit_0\in B(0,R)$, $\rho_{0}^{\prime}>\rho_{0}$ and  $t>T_{0}$ $\left( with \quad T_{0}=\frac{1}{\nu\lambda_{1}}ln\frac{R^2}{{\rho_{0}^{\prime}}^2-{\rho_{0}}^2}\right)$ : 
\begin{equation}
\int_{t}^{t+r}\left\Vert  \wit (s)\right\Vert ^{2}_{1}ds\leqslant\frac{r}{\nu^2\lambda_{1}} \left\Vert \mathbf{f} \right\Vert^{2}+\frac{{\rho_{0}^{\prime}}^2}{\nu}  .
\label{ineq3}%
\end{equation}

$\newline$
{\bf Absorbing set in ${\bf H}_1$: } We take now the inner product of  the first equation of system $(\ref{pb})$ with $A\wit$ to obtain 
\begin{equation}
\frac{1}{2}\frac{d}{dt}\left\Vert \wit\right\Vert ^{2}_{1}+ b(H_{N} (\wit), \wit, A\wit)+\nu|| A\wit ||^{2}=(H_{N}(\mathbf{f}),A\wit) ,
\end{equation}
 leading to
 \begin{equation}
\frac{1}{2}\frac{d}{dt}\left\Vert \wit\right\Vert ^{2}_{1}+ +\nu|| A\wit ||^{2}\leqslant \frac{1}{\nu}\left\Vert H_{N}(\mathbf{f})\right\Vert^{2}+\frac{\nu}{4}\left\Vert A\wit\right\Vert^{2} +|b(H_{N} (\wit), \wit, A\wit)|,
\end{equation}
The trilinear form $b$ satisfies the folowing inequality (see in \cite{LL08}) :%
\begin{equation}
|b(\vit,\mathbf{v},\wit)| \leqslant c^{\prime}\left\Vert \vit \right\Vert^{1/4}\left\Vert \vit \right\Vert^{3/4}_{1} \left\Vert \mathbf{v}\right\Vert^{1/4}_1 \left\Vert A\mathbf{v}\right\Vert^{3/4}\left\Vert \wit\right\Vert .
\end{equation}
Therefore, one has
\begin{equation}
 |b(H_{N} (\wit), \wit, A\wit)| \leqslant c^{\prime}\left\Vert H_{N} (\wit) \right\Vert^{1/4}\left\Vert H_{N} (\wit) \right\Vert^{3/4}_{1} \left\Vert \wit\right\Vert^{1/4}_1 \left\Vert A\wit\right\Vert^{7/4}  .
\end{equation}
Using  $(\ref{H2})$ we have $\left\Vert H_{N}(\wit)\right\Vert_{1}\leqslant \left\Vert H_{N}(\wit)\right\Vert_{2}\leqslant C( \delta,N)\left\Vert \wit\right\Vert$ and using $(\ref{H2})$ :
\begin{equation}
 |b(H_{N} (\wit), \wit, A\wit)| \leqslant C^{\prime}( \delta,N)\left\Vert \wit \right\Vert \left\Vert \wit\right\Vert^{1/4}_1 \left\Vert A\wit\right\Vert^{7/4}  .
\end{equation}
By Young inequality we obtain
\begin{equation}
 |b(H_{N} (\wit), \wit, A\wit)| \leqslant \frac{\nu}{4}\left\Vert A\wit\right\Vert^{2}+\frac{C_1( \delta,N)}{2}\left\Vert \wit \right\Vert^8 \left\Vert \wit\right\Vert^{2}_1 ,
\end{equation}
thus
\begin{equation}
\frac{d}{dt}\left\Vert \wit\right\Vert ^{2}_{1}+\nu|| A\wit ||^{2}\leqslant \frac{2}{\nu}\left\Vert H_{N}(\mathbf{f})\right\Vert^{2}+C_1( \delta,N)\left\Vert \wit \right\Vert^8 \left\Vert \wit\right\Vert^{2}_1
\end{equation}
We now use a Gronwall type proposition (see the proof in \cite{RT88}):  
\begin{proposition}Ê\label {gronwall}
Assume that $y$, $g$ and $h$ are positive, localy integrable functions on $]t_0,+\infty[$, and that for $t \geqslant t_0$, $$\frac{dy}{dt}\leqslant gy+h, \quad\int_{t}^{t+r}y(s)ds\leqslant k_1, \quad\int_{t}^{t+r}g(s)ds,\leqslant k_2,  \quad\int_{t}^{t+r}h(s)ds\leqslant k_3 ,$$
where $r$, $k_1$, $k_2$, $k_3$ are four positive constants, then 
\smallskip
 
\hskip 4cm $\displaystyle y(t+r)\leqslant \left(\frac{k_1}{r}+k_3\right)e^{k_2} , \quad \forall t \geqslant t_0. $  \hfill $\blacksquare$
\end{proposition}

$\newline$
We can now finish the proof. Thanks to $(\ref{estimationH})$ and $(\ref{ineq3})$,  using this lemma with $y= \left\Vert \wit\right\Vert^{2}_1,\quad$ $g=C_1( \delta,N)\left\Vert \wit \right\Vert^8\quad$ and $h=\frac{2}{\nu}\left\Vert H_{N}(\mathbf{f})\right\Vert^{2}$,  we obtain, 
\begin{equation}\label {estimationV}
 \left\Vert \wit (t)\right\Vert^{2}_1\leqslant \left(\frac{k_1}{r}+k_3\right)e^{k_2} , \quad \forall t \geqslant T_0+r,
\end{equation}
with $\quad k_1=\frac{r}{\nu^2\lambda_{1}} \left\Vert \mathbf{f} \right\Vert^{2}+\frac{1}{\nu} {\rho_{0}^{\prime}}^2$, $\quad k_2=C_1( \delta,N){\rho_{0}^{\prime}}^8$, $\quad k_3=\frac{2r}{\nu} \left\Vert \mathbf{f} \right\Vert^{2}$.

Thus, after a time $T_1=T_1\left( \left\Vert \wit_{0}\right\Vert, \left\Vert \mathbf{f} \right\Vert, \nu  \right) $,  $\wit$ is included in a ball or radius $R=R\left( \left\Vert \mathbf{f} \right\Vert, \nu, \delta,N  \right ) $. One deduces that there exists an absorbing set in ${\bf H}_1$.

Let B be a bounded set in ${\bf H}_1$. Estimate $(\ref{estimationV})$ implies that $\displaystyle  {{\bigcup_{t \ge T_0+r} S(t) B}}$ is a bounded set in ${\bf H}_1$ wich is compactly imbeded in ${\bf H}_0$, so $S(t)$ is uniformly compact.
Estimate $(\ref{estimationV})$ also implies  the existence of an absorbing bonded set since $k_1$, $k_2$ and $k_3$ are independant of  $  \wit_{0}$.
Thanks to $(\ref{ATTR})$, this achieves the proof of the theorem. \hfill $\blacksquare$

\bibliographystyle{siam} \bibliography{biblio}

\end{document}